\documentclass[twocolumn,showpacs,preprintnumbers,amsmath,amssymb,superscriptaddress]{revtex4-1}
\usepackage{graphicx}
\usepackage{dcolumn}
\usepackage{bm}
\usepackage{amsmath}

\begin{document}

\title{On the fundamental aspect of the first Kelvin's relation in thermoelectricity}

\author{Y. Apertet}
\affiliation{Lyc\'ee Jacques Pr\'evert, F-27500 Pont-Audemer, France}
\author{C. Goupil}
\affiliation{Laboratoire Interdisciplinaire des Energies de Demain (LIED), UMR 8236 Universit\'e Paris Diderot, CNRS, 5 Rue Thomas Mann, 75013 Paris, France}


\begin{abstract}
Kelvin's relations may be considered as cornerstones in the theory of thermoelectricity. Indeed, they gather together the three thermoelectric effects, associated respectively with Seebeck, Peltier and Thomson, to get a unique and consistent description of thermoelectric phenomena. However, their physical status in literature are quite different. On the one hand, the second Kelvin's relation is associated with the microscopic reversibility, considered as a fundamental thermodynamical property. On the other hand, the first Kelvin's relation is traditionally introduced only as a convenient mathematical relation between Seebeck and Thomson coefficients. In the present article, we stress that, contrary to common believes, this relation may demonstrates deeper insights than a bold mathematical expression between thermoelectric coefficients. It actually reflects the coexistence of two different mechanisms taking place inside thermoelectric systems: Energy conversion and reorganization of heat flow when Seebeck coefficient varies. 
\end{abstract}%


\maketitle

\section{Introduction}

In 1851, Kelvin (while still being known as Thomson) proposed a theoretical framework encompassing both Seebeck and Peltier effects \cite{Thomson1851}, two thermoelectric phenomena then recently discovered \cite{Seebeck1826, Peltier1834}. He also predicted a third effect that, contrary to these two previous effects, does not occur at a junction between different materials. The Thomson effect refers to the evolution of heat as an electric current flows through a temperature gradient in a material: Depending on the current direction, a fraction of heat, sometimes referred to as \emph{Thomson heat}, may be absorb or release along the system. 

In order to evaluate the Thomson heat, Kelvin introduced the Thomson coefficient $\tau$. The volumetric Thomson heat production rate is then given by $\tau {\bf j}\cdot{\bf \nabla}(T)$, with ${\bf j}$ the electrical current density and ${\bf \nabla}(T)$ the local temperature gradient. This additional effect is related to both Peltier and Seebeck effects by the first Kelvin's relation (Eq.~(e) of \cite{Thomson1851}):       
\begin{equation}\label{Kelvin1}
\frac{{\rm d}\Pi}{{\rm d} T} = \alpha + \tau,
\end{equation}
\noindent where $\alpha$ is the Seebeck coefficient and  $\Pi$ is the Peltier coefficient. Note that these same coefficients are linked through the second Kelvin's relation (Eq.~(d) of \cite{Thomson1851}):
\begin{equation}\label{Kelvin2}
\Pi = \alpha T 
\end{equation}
\noindent with $T$ the local temperature. Combining both Kelvin's relations, the Thomson coefficient $\tau$ can be related to the Seebeck coefficient variation with temperature:
\begin{equation}\label{Thomson}
\tau = T \frac{{\rm d}\alpha}{{\rm d} T}.
\end{equation}

Since this seminal work, Kelvin's relations have not been treated on an equal footing. Indeed, the second Kelvin's relation appears now as a specific case of the Onsager's reciprocal relations \cite{Onsager1931, Callen1948}. As such, this relation lies on microscopic reversibility, a fundamental principle in physics. On the contrary, the first Kelvin's relation is believed to demonstrate fewer insights and is often resumed to Eq.~(\ref{Thomson}), i.e., a convenient mathematical expression. We believe however that this difference is not fully justified as Eq.~(\ref{Kelvin1}) also contains interesting physics, in particular regarding the way heat evolves along thermoelectric systems. The article is thus organized as follows. First, in Sec.~\ref{description}, starting from the constitutive local relations for both electrical and heat current densities, we derive the heat equation inside a thermoelectric system, focusing especially on the physical signification of each term. We thus highlight the different thermoelectric mechanisms leading to a modification of the heat current. We then show in Sec.~\ref{interpretation} that the distinction between these mechanisms are actually at the heart of the first Kelvin's relation. We end our discussion with some examples of the practical applications of the Thomson effect. 

\section{\label{description}Description of the system}
\subsection{Constitutive relations}
Thermoelectricity describes the coupled transport of heat and electrical charges. On a local scale, both electrical current density ${\bf j}$ and thermal current density ${\bf q}$ may be related to the thermodynamical forces giving birth to these fluxes, i.e., the gradient of the electrochemical potential ${\bf\nabla} \widetilde{\mu}$ and the gradient of the temperature ${\bf \nabla} T$, using the following phenomenological relations:
\begin{eqnarray}
{\bf j} & = &\sigma \left[{\bf\nabla}(\frac{\widetilde{\mu}}{e}) - \alpha {\bf\nabla}(T) \right],\label{electricaldensity}\\
{\bf q} & = &\alpha T {\bf j} - \kappa {\bf \nabla}(T), \label{heatdensity}
\end{eqnarray} 
\noindent where $\sigma$ is the electrical conductivity,  $\kappa$ is the thermal conductivity and $e$ is the elementary electrical charge (we consider here without loss of generality that charge carriers are electrons). Thermoelectric phenomena appear through  the Seebeck coefficient $\alpha$, reflecting the creation of an electromotive force due to the presence of a thermal gradient, and through the Peltier coefficient $\Pi$, reflecting the transport of heat by each electrical carrier. For clarity, we choose to express the Peltier coefficient in Eq.~(\ref{heatdensity}) as a function of the Seebeck coefficient using the second Kelvin's relation. According to Kelvin, the first term in the right hand side of Eq.~(\ref{heatdensity}) may thus be viewed as a \emph{convective heat current inside the material} since it stems from the global displacement of the electrical carriers \cite{Thomson1856} while the second term may be viewed as \emph{conductive heat current} since it results from a heat propagation without any net carriers movement.

\subsection{Energy conversion}
Due to this coupling between heat and electrical current, thermoelectricity can be used to convert thermal power into electrical power and \emph{vice-versa}. We consider a thermoelectric energy converter with thermally insulated sides and connected with a load (passive if the converter is used as a generator or active if the converter is used as a refrigerator). Our analysis is restrained to steady state. The global energy current density ${\bf w}$ may be decomposed as heat and work contributions and then reads: ${\bf w} = {\bf q} + \widetilde{\mu}{\bf j}_N$, ${\bf j}_N$ being the electron flux, or equivalently  ${\bf w} = {\bf q} - (\widetilde{\mu}/e){\bf j}$.
Due to the steady state condition, and neglecting non-equilibrium carriers, i.e., ${\bf \nabla}\cdot{\bf j} = 0$, the energy conservation reads:
\begin{equation}\label{heat1}
{\bf \nabla}\cdot{\bf q} = {\bf\nabla}(\frac{\widetilde{\mu}}{e}) \cdot {\bf j}.
\end{equation}
\noindent The right hand side of the previous equation is the power delivered from the load to the converter. If the thermodynamical force ${\bf\nabla}(\widetilde{\mu}/e)$ and the electrical current ${\bf j}$ have opposite directions, this term is negative and the converter thus works as a generator, delivering electrical power to the load. Conversely, if they have identical directions, the converter works as a heat pump, using electrical power to impose a heat current. It is possible to decompose the power delivered to  the converter by the load, ${\bf\nabla}(\widetilde{\mu}/e) \cdot {\bf j}$, using Eq.~(\ref{electricaldensity}):
\begin{equation}\label{electromotivepower}
{\bf\nabla}(\frac{\widetilde{\mu}}{e}) \cdot {\bf j} = \alpha {\bf j} {\bf\nabla}(T) + \frac{{\bf j}^2}{\sigma}.
\end{equation}
\noindent While the second term of the right hand side of this relation is easily identified with Joule heating, the interpretation of the first term is less trivial. It has been, for example, mistakenly associated with Thomson effect \cite{Logvinov2006,Titov2015} even if this term has no relation with this effect since it represents the power generated by the Seebeck effect. Indeed, due to Seebeck effect, the thermal gradient generates an electromotive force $ - \alpha {\bf\nabla}(T)$. The term $- \alpha {\bf j} {\bf\nabla}(T)$ is thus the electrical power resulting from this electromotive force and delivered to the load. Unfortunately this power cannot be transfered entirely to the load since, as electrical current gives birth to inevitable losses though Joule heating, a fraction is converted back to thermal energy by the internal electrical resistance of the converter as demonstrated by Eq.~(\ref{electromotivepower}). In practice, these losses are not negligible compared to the produced electrical power: When a thermoelectric generator works at maximum output power, power lost by Joule heating and output power demonstrate identical values, i.e., half of the available electromotive power \cite{Ioffe1960}. 

\subsection{Heat equation}

While Eq.~(\ref{heat1}) is sometimes presented as the heat balance equation (see e.g. \cite{Logvinov2006}), it does not allow to directly obtain the temperature profile inside the converter. To derive a more functional equation, one needs to also consider the definition of the heat current density given by Eq.~(\ref{heatdensity}).
Differentiating that equation with respect to spatial position gives:
\begin{equation}\label{heat2}
{\bf \nabla}\cdot{\bf q} = \alpha {\bf j}\cdot {\bf \nabla}(T) + T {\bf j}\cdot{\bf \nabla}(\alpha)   - {\bf \nabla}\cdot\left(\kappa {\bf \nabla}(T) \right).
\end{equation}
\noindent The combination of this relation with Eqs.~(\ref{heat1}) and (\ref{electromotivepower}) yields the heat equation, sometimes called Domenicali's equation \cite{Domenicali1954}, necessary to obtain the temperature profile inside the converter:
\begin{equation}\label{heatequation}
{\bf \nabla}\cdot\left(\kappa {\bf \nabla}(T) \right) = - \frac{{\bf j}^2}{\sigma} +  T {\bf j}\cdot{\bf \nabla}(\alpha).
\end{equation}
\noindent 
This equation demonstrates that, while the variation of the global heat current density ${\bf q}$ is only related to energy conversion as showed in Eq.~(\ref{heat1}), there is also, at any point of the device, a redistribution of the heat between the two different way of propagation, i.e. the \emph{thermoelectric convection} and the \emph{thermal conduction}. Indeed, the variation of the conductive thermal flux, i.e., $\kappa {\bf \nabla}(T)$, stems from Joule heating but also from spatial variation of the convective thermal flux due to Seebeck coefficient variation.


\section{\label{interpretation}Interpretation of the first Kelvin's relation}

In order to get physical insights from the first Kelvin's relation, we use a viewpoint proposed by Kelvin in Ref.~\cite{Thomson1854}: The Seebeck coefficient is identified with the \emph{specific heat of moving carriers} per unit charge. This vision is consistent with the fact that heat convectively transported by each electrical carrier is given by $-e\alpha T$ as shown in Eq.~(\ref{heatdensity}). Then, an increase (a decrease) of this specific heat leads moving electrical carriers to absorb (release) some heat and, by consequence, to exchange with the conductive part of the thermal current or to absorb Joule heating. This behavior actually corresponds to the heat equation given in Eq.~(\ref{heatequation}). 
The first Kelvin's relation should then be seen as a decomposition of the variation of heat convectively transported by the electrons. As electrical carriers experience a temperature gradient, the convective heat per electrical charge $\Pi$ indeed varies along the system due to two distinct mechanisms: A fraction of this heat is converted into electrochemical energy while another fraction may be absorbed, or released, due to the variation of the specific heat, i.e., the Seebeck coefficient, with temperature. The energy conversion corresponds to the first term of the right hand side of Eq.~(\ref{Kelvin1}) and stems from the variation of temperature at constant Seebeck coefficient. This mechanism represents the useful part of the thermoelectric phenomena. The second mechanism is associated with the second term of the right hand side of Eq.~(\ref{Kelvin1}), i.e., the Thomson coefficient. Note that the associated Thomson effect is restrained in that relation to the variation of the Seebeck coefficient with temperature rather than with spatial position as it naturally appears in Eq.~(\ref{heatequation}). Seebeck coefficient might however vary due to other parameters. If this variation is due to a change of material, the release (or absorption) of heat is the so-called Peltier effect. So, both Peltier and Thomson effects appears to be two manifestations of a same physical phenomenon, i.e., heat exchange due to the variation of the specific heat of moving carriers \cite{Goupil2011}. Despite its obvious pedagogical interest, the identification of Seebeck coefficient with specific heat is seldom found in literature even if a similarity between these two quantities has already been noticed in the case of the free electron gas \cite{Behnia2004}.
 
The contribution of the Thomson effect in the heat equation can be more clearly highlighted using Eq.~(\ref{Thomson}). One thus gets:
\begin{equation}\label{heatequation2}
{\bf \nabla}\cdot\left(\kappa {\bf \nabla}(T) \right) = - \frac{{\bf j}^2}{\sigma} +  \tau {\bf j}\cdot{\bf \nabla}(T).
\end{equation}
\noindent In this form, heat equation becomes independent of any specific design, only depending on the used material. We stress that our derivations, and in particular the previous equation, are valid only if the thermoelectric system has thermally insulated sides, i.e., it exchanges heat only with thermal reservoirs to which it is connected. Obviously, this assumption does not correspond to the practical case envisaged by Kelvin since Thomson effect is experimentally probed by the additional heat escaping through the sides of the system. A more realistic but yet debatable approach consists in considering that temperature distribution remains identical with or without electrical current \cite{Callen1985}: The temperature gradient is thus assumed to be constant in case of a constant thermal conductivity. Consequently, both Joule heating and Thomson heat are then released into the surrounding environment.

\section{Practical applications of the Thomson effect}

As a junction between different materials is mandatory to observe both Peltier and Seebeck effects, the associated coefficients cannot be experimentally determined for a single material contrary to Thomson coefficient. This latter thus appears as a way to evaluate the absolute Seebeck coefficient for a given material at any temperature using Eq.~(\ref{Thomson}) and the fact that in a superconducting state, Seebeck coefficient vanishes (see e.g., \cite{Roberts1977, Martin2010}). 

The Thomson effect has no direct application in energy conversion contrary to the Seebeck effect or the Peltier effect. Indeed, on a local scale, this effect induces a transfer from conductive heat to convective heat (or vice versa) but not a conversion of heat into work. However, it may significantly influence the optimization of thermoelectric systems as it modifies the temperature profile along the device. The impact of the Thomson effect on optimal working conditions of thermoelectric generators and coolers has thus been the subject of numerous studies: See for examples Ref.~\cite{Ybarrondo1965, Snyder2012} for cooling applications and Refs.~\cite{Sunderland1964, Chen1996, Sandoz2013} for electrical power generation.

\section{Conclusions}
In this article, we have discussed the physical meaning of the first Kelvin's relation. We built our analysis on the identification of Seebeck coefficient with the specific capacity of moving carriers, as suggested by Kelvin. We reached the conclusion that the first Kelvin's relation amounts to distinguishing two different mechanisms ruling the evolution of heat convectively transported along the system by the electrical current. Theses two mechanisms are respectively the energy conversion from heat into work and the heat flux reorganization due to variations of specific capacity of electrons, i.e., variations of the Seebeck coefficient. To that extent, we believe that the first Kelvin's relation should be considered as a fundamental and meaningful physical relation rather than just a convenient mathematical expression.


\section*{Ackowledgement}
We are pleased to thank A. Georges for bringing this topic to our attention.


\begin{thebibliography}{12}

\bibitem{Thomson1851} W. Thomson, ``On a mechanical theory of thermo-electric currents'', Proc. R. Soc. Edinb., 91-98 (1851); see also W. Thomson, \emph{Mathematical and Physical Papers} (Cambridge University Press, London, 1882), Art. XLVIII.

\bibitem{Seebeck1826} T. J. Seebeck, ``Ueber die Magnetische Polarisation der Metalle und Erze durch Temperatur-Differenz,'' Annalen der Physik {\bf 82}, 253-286 (1826). 

\bibitem{Peltier1834} J. C. A. Peltier, ``Nouvelles exp\'eriences sur la caloricit\'e des courants \'electrique,'' Annales de Chimie et de Physique {\bf 56}, 371 (1834).


\bibitem{Onsager1931} L. Onsager, ``Reciprocal Relations in Irreversible Processes. I.,'' Phys. Rev. {\bf 37}, 405 (1931).

\bibitem{Callen1948} H. B. Callen, ``The Application of Onsager's Reciprocal Relations to Thermoelectric, Thermomagnetic, and Galvanomagnetic Effects,'' Phys. Rev. {\bf 73}, 1349 (1948).

\bibitem{Thomson1856} W. Thomson, ``The Bakerian Lecture. On the Electrodynamic Qualities of Metals,'' Phil. Trans. R. Soc. Lond. {\bf 146}, 649 (1856); see also W. Thomson, \emph{Mathematical and Physical Papers} (Cambridge University Press, London, 1882), Art. XCI.

\bibitem{Logvinov2006} G. N. Logvinov, J. E. Vel\'azquez, I. M. Lashkevych and Yu. G. Gurevich, ``Heating and cooling in semiconductor structures by an electric current,'' APL {\bf 89}, 092118 (2006).

\bibitem{Titov2015} O. Yu. Titov, J.E. Velazquez-Perez and Yu. G. Gurevich, ``Mechanisms of the thermal electromotive force, heating and cooling in semiconductor structures,''  Int. J. Therm. Sci. {\bf 92},  44-49 (2015).

\bibitem{Ioffe1960} Ioffe A F 1960 \emph{Physics of Semiconductors} (Infosearch, Ltd., London) 


\bibitem{Domenicali1954} C. A. Domenicali, ``Stationary temperature distribution in an electrically heated conductor,'' J. Appl. Phys. {\bf 25}, 1310 (1954).

\bibitem{Thomson1854} W. Thomson, ``Account of Researches in Thermo-Electricity,'' Proc. R. Soc. Lond.  {\bf 7}, 49-58 (1854)  see also W. Thomson, \emph{Mathematical and Physical Papers} (Cambridge University Press, London, 1882), Art. LI.

\bibitem{Goupil2011} C. Goupil, W. Seifert, K. Zabrocki, E. M\"uller, and G. J. Snyder, ``Thermodynamics of Thermoelectric Phenomena and Applications,'' Entropy {\bf 13} 1481-1517 (2011).

\bibitem{Behnia2004} K. Behnia, D. Jaccard, and J. Flouquet, ``On the thermoelectricity of correlated electrons in the zero-temperature limit,'' J. Phys.:Condens. Matter {\bf 16}, 5187 (2004).

\bibitem{Callen1985} H. B. Callen, \emph{Thermodynamics and an introduction to Thermostatistics} (John Wiley \& Sons, New York, 1985).

\bibitem{Roberts1977} R. B. Roberts, ``The absolute scale of thermoelectricity,'' Philosophical Magazine {\bf 36}, 91-107 (1977).

\bibitem{Martin2010} J. Martin, T. Tritt, and C. Uher, ``High temperature Seebeck coefficient metrology,'' J. Appl. Phys. {\bf 108}, 121101 (2010).

\bibitem{Ybarrondo1965} L. J. Ybarrondo and J. E. Sunderland, ``Influence of spatially dependent properties on the performance of a thermoelectric heat pump,'' Advanced Energy Conversion {\bf 5}, 383-405 (1965).

\bibitem{Snyder2012} G. J. Snyder, E. S. Toberer, R. Khanna, and W. Seifert, ``Improved Thermoelectric Cooling Based on the Thomson Effect,'' Phys. Rev. B {\bf 86}, 045202 (2012).

\bibitem{Sunderland1964} J. E. Sunderland and N. T. Burak, ``The influence of the Thomson effect on the performance of a thermoelectric power generator,'' Solid-State Electronics {\bf 7}, 465–471 (1964).
 
\bibitem{Chen1996} J. Chen, Z. Yan, and L. Wu, ``The influence of Thomson effect on the maximum power output and maximum efficiency of a thermoelectric generator,'' J. Appl. Phys. {\bf 79}, 8823 (1996).

\bibitem{Sandoz2013} E. J. Sandoz-Rosado, S. J. Weinstein, and R. J. Stevens, ``On the Thomson effect in thermoelectric power devices,'' Int. J. Therm. Sci. {\bf 66},  1-7 (2013).


\end{thebibliography}
\end{document}